\documentclass[12pt]{article}
\usepackage{epsfig,amsfonts,amssymb}
\usepackage{hyperref}

\usepackage{cite}
\usepackage{cite}

\def\ZZZ{{\hbox{ Z\kern-1.6mm Z}}}
\def\RRR{{\hbox{ R\kern-2.4mm R}}}
\def\CCC{{\hbox{ C\kern-2.0mm C}}}
\def\zzz{{\hbox{z\kern-1mm z}}}

\newcommand{\qeq}{{\hbox{=\kern-2.3mm ? \kern.5mm }}}
\renewcommand{\qeq}{=}

\newcommand{\eps}{\epsilon}

\newcommand{\CC}{{\cal C}}

\newcommand{\OO}{{\cal O}}

\newcommand{\NN}{{\cal N}}

\newcommand{\be}{\begin{equation}}
\newcommand{\ee}{\end{equation}}
\newcommand{\ben}{\begin{eqnarray}\displaystyle}
\newcommand{\een}{\end{eqnarray}}

\newcommand{\refb}[1]{(\ref{#1})}
\newcommand{\p}{\partial}
\newcommand{\sectiono}[1]{\section{#1}\setcounter{equation}{0}}

\def\one{{\hbox{ 1\kern-.8mm l}}}
\def\zero{{\hbox{ 0\kern-1.5mm 0}}}

\newcommand{\bea}[1]{\begin{eqnarray}\label{#1} }
\newcommand{\eea}{\end{eqnarray}}

\newcommand{\eqref}{\refb}

\usepackage{bm}
\usepackage[table]{xcolor}

\def\figone{

\def\JPicScale{0.8}
\ifx\JPicScale\undefined\def\JPicScale{1}\fi
\unitlength \JPicScale mm
\begin{picture}(90,50)(0,0)
\linethickness{.5mm}
\put(30,45){\line(1,0){20}}
\linethickness{0.2mm}
\put(50,45){\line(1,0){20}}
\linethickness{.5mm}
\put(70,45){\line(1,0){20}}

\put(50,45){\makebox(0,0)[cc]{$\times$}}

\put(70,45){\makebox(0,0)[cc]{$\times$}}

\put(40,48){\makebox(0,0)[cc]{$\CC$}}

\put(60,48){\makebox(0,0)[cc]{$\OO$}}

\put(80,48){\makebox(0,0)[cc]{$\CC$}}

\end{picture}

}

\def\figtwo{

\def\JPicScale{0.8}
\ifx\JPicScale\undefined\def\JPicScale{1}\fi
\unitlength \JPicScale mm
\begin{picture}(120,80)(0,0)
\linethickness{.5mm}
\put(20,60){\line(1,0){20}}
\linethickness{0.5mm}
\put(40,60){\line(1,0){20}}
\linethickness{.5mm}
\multiput(80,40)(0.12,0.12){167}{\line(1,0){0.12}}
\linethickness{.5mm}
\multiput(100,60)(0.12,-0.12){167}{\line(1,0){0.12}}
\linethickness{.5mm}
\put(100,60){\line(0,1){20}}
\put(100,80){\makebox(0,0)[cc]{$\times$}}

\put(100.1,59.5){\makebox(0,0)[cc]{$\times$}}

\put(40,60){\makebox(0,0)[cc]{$\times$}}

\put(30,63){\makebox(0,0)[cc]{$\CC$}}

\put(50,63){\makebox(0,0)[cc]{$\CC$}}

\put(93,50){\makebox(0,0)[cc]{$\CC$}}

\put(107,50){\makebox(0,0)[cc]{$\CC$}}

\put(102,70){\makebox(0,0)[cc]{$\CC$}}

\end{picture}

}

\def\figthree{

\def\JPicScale{0.8}
\ifx\JPicScale\undefined\def\JPicScale{1}\fi
\unitlength \JPicScale mm
\begin{picture}(90,50)(0,0)
\linethickness{.5mm}
\put(30,50){\line(1,0){20}}
\linethickness{0.2mm}
\put(50,50){\line(1,0){20}}
\linethickness{.5mm}
\put(70,50){\line(1,0){20}}

\linethickness{0.2mm}
\put(50,30){\line(0,1){20}}

\put(50,50){\makebox(0,0)[cc]{$\times$}}

\put(70,50){\makebox(0,0)[cc]{$\times$}}

\put(40,53){\makebox(0,0)[cc]{$\CC$}}

\put(60,53){\makebox(0,0)[cc]{$\OO$}}

\put(80,53){\makebox(0,0)[cc]{$\CC$}}

\put(52.5,40){\makebox(0,0)[cc]{$\OO$}}

\end{picture}

}

\def\figfour{

\def\JPicScale{0.8}
\ifx\JPicScale\undefined\def\JPicScale{1}\fi
\unitlength \JPicScale mm
\begin{picture}(80.33,60.33)(0,0)
\linethickness{0.2mm}
\put(80.32,44.75){\line(0,1){0.5}}
\multiput(80.31,45.75)(0.02,-0.5){1}{\line(0,-1){0.5}}
\multiput(80.27,46.25)(0.03,-0.5){1}{\line(0,-1){0.5}}
\multiput(80.22,46.75)(0.05,-0.5){1}{\line(0,-1){0.5}}
\multiput(80.16,47.25)(0.07,-0.5){1}{\line(0,-1){0.5}}
\multiput(80.08,47.74)(0.08,-0.49){1}{\line(0,-1){0.49}}
\multiput(79.98,48.24)(0.1,-0.49){1}{\line(0,-1){0.49}}
\multiput(79.87,48.72)(0.11,-0.49){1}{\line(0,-1){0.49}}
\multiput(79.74,49.21)(0.13,-0.48){1}{\line(0,-1){0.48}}
\multiput(79.59,49.69)(0.15,-0.48){1}{\line(0,-1){0.48}}
\multiput(79.43,50.16)(0.16,-0.47){1}{\line(0,-1){0.47}}
\multiput(79.25,50.63)(0.18,-0.47){1}{\line(0,-1){0.47}}
\multiput(79.06,51.1)(0.1,-0.23){2}{\line(0,-1){0.23}}
\multiput(78.85,51.55)(0.1,-0.23){2}{\line(0,-1){0.23}}
\multiput(78.63,52)(0.11,-0.22){2}{\line(0,-1){0.22}}
\multiput(78.4,52.44)(0.12,-0.22){2}{\line(0,-1){0.22}}
\multiput(78.14,52.88)(0.13,-0.22){2}{\line(0,-1){0.22}}
\multiput(77.88,53.3)(0.13,-0.21){2}{\line(0,-1){0.21}}
\multiput(77.6,53.72)(0.14,-0.21){2}{\line(0,-1){0.21}}
\multiput(77.31,54.13)(0.15,-0.2){2}{\line(0,-1){0.2}}
\multiput(77,54.53)(0.1,-0.13){3}{\line(0,-1){0.13}}
\multiput(76.69,54.91)(0.11,-0.13){3}{\line(0,-1){0.13}}
\multiput(76.36,55.29)(0.11,-0.13){3}{\line(0,-1){0.13}}
\multiput(76.01,55.66)(0.11,-0.12){3}{\line(0,-1){0.12}}
\multiput(75.66,56.01)(0.12,-0.12){3}{\line(0,-1){0.12}}
\multiput(75.29,56.36)(0.12,-0.11){3}{\line(1,0){0.12}}
\multiput(74.91,56.69)(0.13,-0.11){3}{\line(1,0){0.13}}
\multiput(74.53,57)(0.13,-0.11){3}{\line(1,0){0.13}}
\multiput(74.13,57.31)(0.13,-0.1){3}{\line(1,0){0.13}}
\multiput(73.72,57.6)(0.2,-0.15){2}{\line(1,0){0.2}}
\multiput(73.3,57.88)(0.21,-0.14){2}{\line(1,0){0.21}}
\multiput(72.88,58.14)(0.21,-0.13){2}{\line(1,0){0.21}}
\multiput(72.44,58.4)(0.22,-0.13){2}{\line(1,0){0.22}}
\multiput(72,58.63)(0.22,-0.12){2}{\line(1,0){0.22}}
\multiput(71.55,58.85)(0.22,-0.11){2}{\line(1,0){0.22}}
\multiput(71.1,59.06)(0.23,-0.1){2}{\line(1,0){0.23}}
\multiput(70.63,59.25)(0.23,-0.1){2}{\line(1,0){0.23}}
\multiput(70.16,59.43)(0.47,-0.18){1}{\line(1,0){0.47}}
\multiput(69.69,59.59)(0.47,-0.16){1}{\line(1,0){0.47}}
\multiput(69.21,59.74)(0.48,-0.15){1}{\line(1,0){0.48}}
\multiput(68.72,59.87)(0.48,-0.13){1}{\line(1,0){0.48}}
\multiput(68.24,59.98)(0.49,-0.11){1}{\line(1,0){0.49}}
\multiput(67.74,60.08)(0.49,-0.1){1}{\line(1,0){0.49}}
\multiput(67.25,60.16)(0.49,-0.08){1}{\line(1,0){0.49}}
\multiput(66.75,60.22)(0.5,-0.07){1}{\line(1,0){0.5}}
\multiput(66.25,60.27)(0.5,-0.05){1}{\line(1,0){0.5}}
\multiput(65.75,60.31)(0.5,-0.03){1}{\line(1,0){0.5}}
\multiput(65.25,60.32)(0.5,-0.02){1}{\line(1,0){0.5}}
\put(64.75,60.32){\line(1,0){0.5}}
\multiput(64.25,60.31)(0.5,0.02){1}{\line(1,0){0.5}}
\multiput(63.75,60.27)(0.5,0.03){1}{\line(1,0){0.5}}
\multiput(63.25,60.22)(0.5,0.05){1}{\line(1,0){0.5}}
\multiput(62.75,60.16)(0.5,0.07){1}{\line(1,0){0.5}}
\multiput(62.26,60.08)(0.49,0.08){1}{\line(1,0){0.49}}
\multiput(61.76,59.98)(0.49,0.1){1}{\line(1,0){0.49}}
\multiput(61.28,59.87)(0.49,0.11){1}{\line(1,0){0.49}}
\multiput(60.79,59.74)(0.48,0.13){1}{\line(1,0){0.48}}
\multiput(60.31,59.59)(0.48,0.15){1}{\line(1,0){0.48}}
\multiput(59.84,59.43)(0.47,0.16){1}{\line(1,0){0.47}}
\multiput(59.37,59.25)(0.47,0.18){1}{\line(1,0){0.47}}
\multiput(58.9,59.06)(0.23,0.1){2}{\line(1,0){0.23}}
\multiput(58.45,58.85)(0.23,0.1){2}{\line(1,0){0.23}}
\multiput(58,58.63)(0.22,0.11){2}{\line(1,0){0.22}}
\multiput(57.56,58.4)(0.22,0.12){2}{\line(1,0){0.22}}
\multiput(57.12,58.14)(0.22,0.13){2}{\line(1,0){0.22}}
\multiput(56.7,57.88)(0.21,0.13){2}{\line(1,0){0.21}}
\multiput(56.28,57.6)(0.21,0.14){2}{\line(1,0){0.21}}
\multiput(55.87,57.31)(0.2,0.15){2}{\line(1,0){0.2}}
\multiput(55.47,57)(0.13,0.1){3}{\line(1,0){0.13}}
\multiput(55.09,56.69)(0.13,0.11){3}{\line(1,0){0.13}}
\multiput(54.71,56.36)(0.13,0.11){3}{\line(1,0){0.13}}
\multiput(54.34,56.01)(0.12,0.11){3}{\line(1,0){0.12}}
\multiput(53.99,55.66)(0.12,0.12){3}{\line(1,0){0.12}}
\multiput(53.64,55.29)(0.11,0.12){3}{\line(0,1){0.12}}
\multiput(53.31,54.91)(0.11,0.13){3}{\line(0,1){0.13}}
\multiput(53,54.53)(0.11,0.13){3}{\line(0,1){0.13}}
\multiput(52.69,54.13)(0.1,0.13){3}{\line(0,1){0.13}}
\multiput(52.4,53.72)(0.15,0.2){2}{\line(0,1){0.2}}
\multiput(52.12,53.3)(0.14,0.21){2}{\line(0,1){0.21}}
\multiput(51.86,52.88)(0.13,0.21){2}{\line(0,1){0.21}}
\multiput(51.6,52.44)(0.13,0.22){2}{\line(0,1){0.22}}
\multiput(51.37,52)(0.12,0.22){2}{\line(0,1){0.22}}
\multiput(51.15,51.55)(0.11,0.22){2}{\line(0,1){0.22}}
\multiput(50.94,51.1)(0.1,0.23){2}{\line(0,1){0.23}}
\multiput(50.75,50.63)(0.1,0.23){2}{\line(0,1){0.23}}
\multiput(50.57,50.16)(0.18,0.47){1}{\line(0,1){0.47}}
\multiput(50.41,49.69)(0.16,0.47){1}{\line(0,1){0.47}}
\multiput(50.26,49.21)(0.15,0.48){1}{\line(0,1){0.48}}
\multiput(50.13,48.72)(0.13,0.48){1}{\line(0,1){0.48}}
\multiput(50.02,48.24)(0.11,0.49){1}{\line(0,1){0.49}}
\multiput(49.92,47.74)(0.1,0.49){1}{\line(0,1){0.49}}
\multiput(49.84,47.25)(0.08,0.49){1}{\line(0,1){0.49}}
\multiput(49.78,46.75)(0.07,0.5){1}{\line(0,1){0.5}}
\multiput(49.73,46.25)(0.05,0.5){1}{\line(0,1){0.5}}
\multiput(49.69,45.75)(0.03,0.5){1}{\line(0,1){0.5}}
\multiput(49.68,45.25)(0.02,0.5){1}{\line(0,1){0.5}}
\put(49.68,44.75){\line(0,1){0.5}}
\multiput(49.68,44.75)(0.02,-0.5){1}{\line(0,-1){0.5}}
\multiput(49.69,44.25)(0.03,-0.5){1}{\line(0,-1){0.5}}
\multiput(49.73,43.75)(0.05,-0.5){1}{\line(0,-1){0.5}}
\multiput(49.78,43.25)(0.07,-0.5){1}{\line(0,-1){0.5}}
\multiput(49.84,42.75)(0.08,-0.49){1}{\line(0,-1){0.49}}
\multiput(49.92,42.26)(0.1,-0.49){1}{\line(0,-1){0.49}}
\multiput(50.02,41.76)(0.11,-0.49){1}{\line(0,-1){0.49}}
\multiput(50.13,41.28)(0.13,-0.48){1}{\line(0,-1){0.48}}
\multiput(50.26,40.79)(0.15,-0.48){1}{\line(0,-1){0.48}}
\multiput(50.41,40.31)(0.16,-0.47){1}{\line(0,-1){0.47}}
\multiput(50.57,39.84)(0.18,-0.47){1}{\line(0,-1){0.47}}
\multiput(50.75,39.37)(0.1,-0.23){2}{\line(0,-1){0.23}}
\multiput(50.94,38.9)(0.1,-0.23){2}{\line(0,-1){0.23}}
\multiput(51.15,38.45)(0.11,-0.22){2}{\line(0,-1){0.22}}
\multiput(51.37,38)(0.12,-0.22){2}{\line(0,-1){0.22}}
\multiput(51.6,37.56)(0.13,-0.22){2}{\line(0,-1){0.22}}
\multiput(51.86,37.12)(0.13,-0.21){2}{\line(0,-1){0.21}}
\multiput(52.12,36.7)(0.14,-0.21){2}{\line(0,-1){0.21}}
\multiput(52.4,36.28)(0.15,-0.2){2}{\line(0,-1){0.2}}
\multiput(52.69,35.87)(0.1,-0.13){3}{\line(0,-1){0.13}}
\multiput(53,35.47)(0.11,-0.13){3}{\line(0,-1){0.13}}
\multiput(53.31,35.09)(0.11,-0.13){3}{\line(0,-1){0.13}}
\multiput(53.64,34.71)(0.11,-0.12){3}{\line(0,-1){0.12}}
\multiput(53.99,34.34)(0.12,-0.12){3}{\line(0,-1){0.12}}
\multiput(54.34,33.99)(0.12,-0.11){3}{\line(1,0){0.12}}
\multiput(54.71,33.64)(0.13,-0.11){3}{\line(1,0){0.13}}
\multiput(55.09,33.31)(0.13,-0.11){3}{\line(1,0){0.13}}
\multiput(55.47,33)(0.13,-0.1){3}{\line(1,0){0.13}}
\multiput(55.87,32.69)(0.2,-0.15){2}{\line(1,0){0.2}}
\multiput(56.28,32.4)(0.21,-0.14){2}{\line(1,0){0.21}}
\multiput(56.7,32.12)(0.21,-0.13){2}{\line(1,0){0.21}}
\multiput(57.12,31.86)(0.22,-0.13){2}{\line(1,0){0.22}}
\multiput(57.56,31.6)(0.22,-0.12){2}{\line(1,0){0.22}}
\multiput(58,31.37)(0.22,-0.11){2}{\line(1,0){0.22}}
\multiput(58.45,31.15)(0.23,-0.1){2}{\line(1,0){0.23}}
\multiput(58.9,30.94)(0.23,-0.1){2}{\line(1,0){0.23}}
\multiput(59.37,30.75)(0.47,-0.18){1}{\line(1,0){0.47}}
\multiput(59.84,30.57)(0.47,-0.16){1}{\line(1,0){0.47}}
\multiput(60.31,30.41)(0.48,-0.15){1}{\line(1,0){0.48}}
\multiput(60.79,30.26)(0.48,-0.13){1}{\line(1,0){0.48}}
\multiput(61.28,30.13)(0.49,-0.11){1}{\line(1,0){0.49}}
\multiput(61.76,30.02)(0.49,-0.1){1}{\line(1,0){0.49}}
\multiput(62.26,29.92)(0.49,-0.08){1}{\line(1,0){0.49}}
\multiput(62.75,29.84)(0.5,-0.07){1}{\line(1,0){0.5}}
\multiput(63.25,29.78)(0.5,-0.05){1}{\line(1,0){0.5}}
\multiput(63.75,29.73)(0.5,-0.03){1}{\line(1,0){0.5}}
\multiput(64.25,29.69)(0.5,-0.02){1}{\line(1,0){0.5}}
\put(64.75,29.68){\line(1,0){0.5}}
\multiput(65.25,29.68)(0.5,0.02){1}{\line(1,0){0.5}}
\multiput(65.75,29.69)(0.5,0.03){1}{\line(1,0){0.5}}
\multiput(66.25,29.73)(0.5,0.05){1}{\line(1,0){0.5}}
\multiput(66.75,29.78)(0.5,0.07){1}{\line(1,0){0.5}}
\multiput(67.25,29.84)(0.49,0.08){1}{\line(1,0){0.49}}
\multiput(67.74,29.92)(0.49,0.1){1}{\line(1,0){0.49}}
\multiput(68.24,30.02)(0.49,0.11){1}{\line(1,0){0.49}}
\multiput(68.72,30.13)(0.48,0.13){1}{\line(1,0){0.48}}
\multiput(69.21,30.26)(0.48,0.15){1}{\line(1,0){0.48}}
\multiput(69.69,30.41)(0.47,0.16){1}{\line(1,0){0.47}}
\multiput(70.16,30.57)(0.47,0.18){1}{\line(1,0){0.47}}
\multiput(70.63,30.75)(0.23,0.1){2}{\line(1,0){0.23}}
\multiput(71.1,30.94)(0.23,0.1){2}{\line(1,0){0.23}}
\multiput(71.55,31.15)(0.22,0.11){2}{\line(1,0){0.22}}
\multiput(72,31.37)(0.22,0.12){2}{\line(1,0){0.22}}
\multiput(72.44,31.6)(0.22,0.13){2}{\line(1,0){0.22}}
\multiput(72.88,31.86)(0.21,0.13){2}{\line(1,0){0.21}}
\multiput(73.3,32.12)(0.21,0.14){2}{\line(1,0){0.21}}
\multiput(73.72,32.4)(0.2,0.15){2}{\line(1,0){0.2}}
\multiput(74.13,32.69)(0.13,0.1){3}{\line(1,0){0.13}}
\multiput(74.53,33)(0.13,0.11){3}{\line(1,0){0.13}}
\multiput(74.91,33.31)(0.13,0.11){3}{\line(1,0){0.13}}
\multiput(75.29,33.64)(0.12,0.11){3}{\line(1,0){0.12}}
\multiput(75.66,33.99)(0.12,0.12){3}{\line(0,1){0.12}}
\multiput(76.01,34.34)(0.11,0.12){3}{\line(0,1){0.12}}
\multiput(76.36,34.71)(0.11,0.13){3}{\line(0,1){0.13}}
\multiput(76.69,35.09)(0.11,0.13){3}{\line(0,1){0.13}}
\multiput(77,35.47)(0.1,0.13){3}{\line(0,1){0.13}}
\multiput(77.31,35.87)(0.15,0.2){2}{\line(0,1){0.2}}
\multiput(77.6,36.28)(0.14,0.21){2}{\line(0,1){0.21}}
\multiput(77.88,36.7)(0.13,0.21){2}{\line(0,1){0.21}}
\multiput(78.14,37.12)(0.13,0.22){2}{\line(0,1){0.22}}
\multiput(78.4,37.56)(0.12,0.22){2}{\line(0,1){0.22}}
\multiput(78.63,38)(0.11,0.22){2}{\line(0,1){0.22}}
\multiput(78.85,38.45)(0.1,0.23){2}{\line(0,1){0.23}}
\multiput(79.06,38.9)(0.1,0.23){2}{\line(0,1){0.23}}
\multiput(79.25,39.37)(0.18,0.47){1}{\line(0,1){0.47}}
\multiput(79.43,39.84)(0.16,0.47){1}{\line(0,1){0.47}}
\multiput(79.59,40.31)(0.15,0.48){1}{\line(0,1){0.48}}
\multiput(79.74,40.79)(0.13,0.48){1}{\line(0,1){0.48}}
\multiput(79.87,41.28)(0.11,0.49){1}{\line(0,1){0.49}}
\multiput(79.98,41.76)(0.1,0.49){1}{\line(0,1){0.49}}
\multiput(80.08,42.26)(0.08,0.49){1}{\line(0,1){0.49}}
\multiput(80.16,42.75)(0.07,0.5){1}{\line(0,1){0.5}}
\multiput(80.22,43.25)(0.05,0.5){1}{\line(0,1){0.5}}
\multiput(80.27,43.75)(0.03,0.5){1}{\line(0,1){0.5}}
\multiput(80.31,44.25)(0.02,0.5){1}{\line(0,1){0.5}}

\linethickness{0.5mm}
\put(65,10){\line(0,1){20}}
\put(65,30){\makebox(0,0)[cc]{$\times$}}

\put(83.5,45){\makebox(0,0)[cc]{$\OO$}}

\put(67.5,20){\makebox(0,0)[cc]{$\CC$}}

\end{picture}

}

\def\figfive{

\def\JPicScale{0.8}
\ifx\JPicScale\undefined\def\JPicScale{1}\fi
\unitlength \JPicScale mm
\begin{picture}(80.33,60.33)(0,0)
\linethickness{0.3mm}
\put(80.32,44.75){\line(0,1){0.5}}
\multiput(80.31,45.75)(0.02,-0.5){1}{\line(0,-1){0.5}}
\multiput(80.27,46.25)(0.03,-0.5){1}{\line(0,-1){0.5}}
\multiput(80.22,46.75)(0.05,-0.5){1}{\line(0,-1){0.5}}
\multiput(80.16,47.25)(0.07,-0.5){1}{\line(0,-1){0.5}}
\multiput(80.08,47.74)(0.08,-0.49){1}{\line(0,-1){0.49}}
\multiput(79.98,48.24)(0.1,-0.49){1}{\line(0,-1){0.49}}
\multiput(79.87,48.72)(0.11,-0.49){1}{\line(0,-1){0.49}}
\multiput(79.74,49.21)(0.13,-0.48){1}{\line(0,-1){0.48}}
\multiput(79.59,49.69)(0.15,-0.48){1}{\line(0,-1){0.48}}
\multiput(79.43,50.16)(0.16,-0.47){1}{\line(0,-1){0.47}}
\multiput(79.25,50.63)(0.18,-0.47){1}{\line(0,-1){0.47}}
\multiput(79.06,51.1)(0.1,-0.23){2}{\line(0,-1){0.23}}
\multiput(78.85,51.55)(0.1,-0.23){2}{\line(0,-1){0.23}}
\multiput(78.63,52)(0.11,-0.22){2}{\line(0,-1){0.22}}
\multiput(78.4,52.44)(0.12,-0.22){2}{\line(0,-1){0.22}}
\multiput(78.14,52.88)(0.13,-0.22){2}{\line(0,-1){0.22}}
\multiput(77.88,53.3)(0.13,-0.21){2}{\line(0,-1){0.21}}
\multiput(77.6,53.72)(0.14,-0.21){2}{\line(0,-1){0.21}}
\multiput(77.31,54.13)(0.15,-0.2){2}{\line(0,-1){0.2}}
\multiput(77,54.53)(0.1,-0.13){3}{\line(0,-1){0.13}}
\multiput(76.69,54.91)(0.11,-0.13){3}{\line(0,-1){0.13}}
\multiput(76.36,55.29)(0.11,-0.13){3}{\line(0,-1){0.13}}
\multiput(76.01,55.66)(0.11,-0.12){3}{\line(0,-1){0.12}}
\multiput(75.66,56.01)(0.12,-0.12){3}{\line(0,-1){0.12}}
\multiput(75.29,56.36)(0.12,-0.11){3}{\line(1,0){0.12}}
\multiput(74.91,56.69)(0.13,-0.11){3}{\line(1,0){0.13}}
\multiput(74.53,57)(0.13,-0.11){3}{\line(1,0){0.13}}
\multiput(74.13,57.31)(0.13,-0.1){3}{\line(1,0){0.13}}
\multiput(73.72,57.6)(0.2,-0.15){2}{\line(1,0){0.2}}
\multiput(73.3,57.88)(0.21,-0.14){2}{\line(1,0){0.21}}
\multiput(72.88,58.14)(0.21,-0.13){2}{\line(1,0){0.21}}
\multiput(72.44,58.4)(0.22,-0.13){2}{\line(1,0){0.22}}
\multiput(72,58.63)(0.22,-0.12){2}{\line(1,0){0.22}}
\multiput(71.55,58.85)(0.22,-0.11){2}{\line(1,0){0.22}}
\multiput(71.1,59.06)(0.23,-0.1){2}{\line(1,0){0.23}}
\multiput(70.63,59.25)(0.23,-0.1){2}{\line(1,0){0.23}}
\multiput(70.16,59.43)(0.47,-0.18){1}{\line(1,0){0.47}}
\multiput(69.69,59.59)(0.47,-0.16){1}{\line(1,0){0.47}}
\multiput(69.21,59.74)(0.48,-0.15){1}{\line(1,0){0.48}}
\multiput(68.72,59.87)(0.48,-0.13){1}{\line(1,0){0.48}}
\multiput(68.24,59.98)(0.49,-0.11){1}{\line(1,0){0.49}}
\multiput(67.74,60.08)(0.49,-0.1){1}{\line(1,0){0.49}}
\multiput(67.25,60.16)(0.49,-0.08){1}{\line(1,0){0.49}}
\multiput(66.75,60.22)(0.5,-0.07){1}{\line(1,0){0.5}}
\multiput(66.25,60.27)(0.5,-0.05){1}{\line(1,0){0.5}}
\multiput(65.75,60.31)(0.5,-0.03){1}{\line(1,0){0.5}}
\multiput(65.25,60.32)(0.5,-0.02){1}{\line(1,0){0.5}}
\put(64.75,60.32){\line(1,0){0.5}}
\multiput(64.25,60.31)(0.5,0.02){1}{\line(1,0){0.5}}
\multiput(63.75,60.27)(0.5,0.03){1}{\line(1,0){0.5}}
\multiput(63.25,60.22)(0.5,0.05){1}{\line(1,0){0.5}}
\multiput(62.75,60.16)(0.5,0.07){1}{\line(1,0){0.5}}
\multiput(62.26,60.08)(0.49,0.08){1}{\line(1,0){0.49}}
\multiput(61.76,59.98)(0.49,0.1){1}{\line(1,0){0.49}}
\multiput(61.28,59.87)(0.49,0.11){1}{\line(1,0){0.49}}
\multiput(60.79,59.74)(0.48,0.13){1}{\line(1,0){0.48}}
\multiput(60.31,59.59)(0.48,0.15){1}{\line(1,0){0.48}}
\multiput(59.84,59.43)(0.47,0.16){1}{\line(1,0){0.47}}
\multiput(59.37,59.25)(0.47,0.18){1}{\line(1,0){0.47}}
\multiput(58.9,59.06)(0.23,0.1){2}{\line(1,0){0.23}}
\multiput(58.45,58.85)(0.23,0.1){2}{\line(1,0){0.23}}
\multiput(58,58.63)(0.22,0.11){2}{\line(1,0){0.22}}
\multiput(57.56,58.4)(0.22,0.12){2}{\line(1,0){0.22}}
\multiput(57.12,58.14)(0.22,0.13){2}{\line(1,0){0.22}}
\multiput(56.7,57.88)(0.21,0.13){2}{\line(1,0){0.21}}
\multiput(56.28,57.6)(0.21,0.14){2}{\line(1,0){0.21}}
\multiput(55.87,57.31)(0.2,0.15){2}{\line(1,0){0.2}}
\multiput(55.47,57)(0.13,0.1){3}{\line(1,0){0.13}}
\multiput(55.09,56.69)(0.13,0.11){3}{\line(1,0){0.13}}
\multiput(54.71,56.36)(0.13,0.11){3}{\line(1,0){0.13}}
\multiput(54.34,56.01)(0.12,0.11){3}{\line(1,0){0.12}}
\multiput(53.99,55.66)(0.12,0.12){3}{\line(1,0){0.12}}
\multiput(53.64,55.29)(0.11,0.12){3}{\line(0,1){0.12}}
\multiput(53.31,54.91)(0.11,0.13){3}{\line(0,1){0.13}}
\multiput(53,54.53)(0.11,0.13){3}{\line(0,1){0.13}}
\multiput(52.69,54.13)(0.1,0.13){3}{\line(0,1){0.13}}
\multiput(52.4,53.72)(0.15,0.2){2}{\line(0,1){0.2}}
\multiput(52.12,53.3)(0.14,0.21){2}{\line(0,1){0.21}}
\multiput(51.86,52.88)(0.13,0.21){2}{\line(0,1){0.21}}
\multiput(51.6,52.44)(0.13,0.22){2}{\line(0,1){0.22}}
\multiput(51.37,52)(0.12,0.22){2}{\line(0,1){0.22}}
\multiput(51.15,51.55)(0.11,0.22){2}{\line(0,1){0.22}}
\multiput(50.94,51.1)(0.1,0.23){2}{\line(0,1){0.23}}
\multiput(50.75,50.63)(0.1,0.23){2}{\line(0,1){0.23}}
\multiput(50.57,50.16)(0.18,0.47){1}{\line(0,1){0.47}}
\multiput(50.41,49.69)(0.16,0.47){1}{\line(0,1){0.47}}
\multiput(50.26,49.21)(0.15,0.48){1}{\line(0,1){0.48}}
\multiput(50.13,48.72)(0.13,0.48){1}{\line(0,1){0.48}}
\multiput(50.02,48.24)(0.11,0.49){1}{\line(0,1){0.49}}
\multiput(49.92,47.74)(0.1,0.49){1}{\line(0,1){0.49}}
\multiput(49.84,47.25)(0.08,0.49){1}{\line(0,1){0.49}}
\multiput(49.78,46.75)(0.07,0.5){1}{\line(0,1){0.5}}
\multiput(49.73,46.25)(0.05,0.5){1}{\line(0,1){0.5}}
\multiput(49.69,45.75)(0.03,0.5){1}{\line(0,1){0.5}}
\multiput(49.68,45.25)(0.02,0.5){1}{\line(0,1){0.5}}
\put(49.68,44.75){\line(0,1){0.5}}
\multiput(49.68,44.75)(0.02,-0.5){1}{\line(0,-1){0.5}}
\multiput(49.69,44.25)(0.03,-0.5){1}{\line(0,-1){0.5}}
\multiput(49.73,43.75)(0.05,-0.5){1}{\line(0,-1){0.5}}
\multiput(49.78,43.25)(0.07,-0.5){1}{\line(0,-1){0.5}}
\multiput(49.84,42.75)(0.08,-0.49){1}{\line(0,-1){0.49}}
\multiput(49.92,42.26)(0.1,-0.49){1}{\line(0,-1){0.49}}
\multiput(50.02,41.76)(0.11,-0.49){1}{\line(0,-1){0.49}}
\multiput(50.13,41.28)(0.13,-0.48){1}{\line(0,-1){0.48}}
\multiput(50.26,40.79)(0.15,-0.48){1}{\line(0,-1){0.48}}
\multiput(50.41,40.31)(0.16,-0.47){1}{\line(0,-1){0.47}}
\multiput(50.57,39.84)(0.18,-0.47){1}{\line(0,-1){0.47}}
\multiput(50.75,39.37)(0.1,-0.23){2}{\line(0,-1){0.23}}
\multiput(50.94,38.9)(0.1,-0.23){2}{\line(0,-1){0.23}}
\multiput(51.15,38.45)(0.11,-0.22){2}{\line(0,-1){0.22}}
\multiput(51.37,38)(0.12,-0.22){2}{\line(0,-1){0.22}}
\multiput(51.6,37.56)(0.13,-0.22){2}{\line(0,-1){0.22}}
\multiput(51.86,37.12)(0.13,-0.21){2}{\line(0,-1){0.21}}
\multiput(52.12,36.7)(0.14,-0.21){2}{\line(0,-1){0.21}}
\multiput(52.4,36.28)(0.15,-0.2){2}{\line(0,-1){0.2}}
\multiput(52.69,35.87)(0.1,-0.13){3}{\line(0,-1){0.13}}
\multiput(53,35.47)(0.11,-0.13){3}{\line(0,-1){0.13}}
\multiput(53.31,35.09)(0.11,-0.13){3}{\line(0,-1){0.13}}
\multiput(53.64,34.71)(0.11,-0.12){3}{\line(0,-1){0.12}}
\multiput(53.99,34.34)(0.12,-0.12){3}{\line(0,-1){0.12}}
\multiput(54.34,33.99)(0.12,-0.11){3}{\line(1,0){0.12}}
\multiput(54.71,33.64)(0.13,-0.11){3}{\line(1,0){0.13}}
\multiput(55.09,33.31)(0.13,-0.11){3}{\line(1,0){0.13}}
\multiput(55.47,33)(0.13,-0.1){3}{\line(1,0){0.13}}
\multiput(55.87,32.69)(0.2,-0.15){2}{\line(1,0){0.2}}
\multiput(56.28,32.4)(0.21,-0.14){2}{\line(1,0){0.21}}
\multiput(56.7,32.12)(0.21,-0.13){2}{\line(1,0){0.21}}
\multiput(57.12,31.86)(0.22,-0.13){2}{\line(1,0){0.22}}
\multiput(57.56,31.6)(0.22,-0.12){2}{\line(1,0){0.22}}
\multiput(58,31.37)(0.22,-0.11){2}{\line(1,0){0.22}}
\multiput(58.45,31.15)(0.23,-0.1){2}{\line(1,0){0.23}}
\multiput(58.9,30.94)(0.23,-0.1){2}{\line(1,0){0.23}}
\multiput(59.37,30.75)(0.47,-0.18){1}{\line(1,0){0.47}}
\multiput(59.84,30.57)(0.47,-0.16){1}{\line(1,0){0.47}}
\multiput(60.31,30.41)(0.48,-0.15){1}{\line(1,0){0.48}}
\multiput(60.79,30.26)(0.48,-0.13){1}{\line(1,0){0.48}}
\multiput(61.28,30.13)(0.49,-0.11){1}{\line(1,0){0.49}}
\multiput(61.76,30.02)(0.49,-0.1){1}{\line(1,0){0.49}}
\multiput(62.26,29.92)(0.49,-0.08){1}{\line(1,0){0.49}}
\multiput(62.75,29.84)(0.5,-0.07){1}{\line(1,0){0.5}}
\multiput(63.25,29.78)(0.5,-0.05){1}{\line(1,0){0.5}}
\multiput(63.75,29.73)(0.5,-0.03){1}{\line(1,0){0.5}}
\multiput(64.25,29.69)(0.5,-0.02){1}{\line(1,0){0.5}}
\put(64.75,29.68){\line(1,0){0.5}}
\multiput(65.25,29.68)(0.5,0.02){1}{\line(1,0){0.5}}
\multiput(65.75,29.69)(0.5,0.03){1}{\line(1,0){0.5}}
\multiput(66.25,29.73)(0.5,0.05){1}{\line(1,0){0.5}}
\multiput(66.75,29.78)(0.5,0.07){1}{\line(1,0){0.5}}
\multiput(67.25,29.84)(0.49,0.08){1}{\line(1,0){0.49}}
\multiput(67.74,29.92)(0.49,0.1){1}{\line(1,0){0.49}}
\multiput(68.24,30.02)(0.49,0.11){1}{\line(1,0){0.49}}
\multiput(68.72,30.13)(0.48,0.13){1}{\line(1,0){0.48}}
\multiput(69.21,30.26)(0.48,0.15){1}{\line(1,0){0.48}}
\multiput(69.69,30.41)(0.47,0.16){1}{\line(1,0){0.47}}
\multiput(70.16,30.57)(0.47,0.18){1}{\line(1,0){0.47}}
\multiput(70.63,30.75)(0.23,0.1){2}{\line(1,0){0.23}}
\multiput(71.1,30.94)(0.23,0.1){2}{\line(1,0){0.23}}
\multiput(71.55,31.15)(0.22,0.11){2}{\line(1,0){0.22}}
\multiput(72,31.37)(0.22,0.12){2}{\line(1,0){0.22}}
\multiput(72.44,31.6)(0.22,0.13){2}{\line(1,0){0.22}}
\multiput(72.88,31.86)(0.21,0.13){2}{\line(1,0){0.21}}
\multiput(73.3,32.12)(0.21,0.14){2}{\line(1,0){0.21}}
\multiput(73.72,32.4)(0.2,0.15){2}{\line(1,0){0.2}}
\multiput(74.13,32.69)(0.13,0.1){3}{\line(1,0){0.13}}
\multiput(74.53,33)(0.13,0.11){3}{\line(1,0){0.13}}
\multiput(74.91,33.31)(0.13,0.11){3}{\line(1,0){0.13}}
\multiput(75.29,33.64)(0.12,0.11){3}{\line(1,0){0.12}}
\multiput(75.66,33.99)(0.12,0.12){3}{\line(0,1){0.12}}
\multiput(76.01,34.34)(0.11,0.12){3}{\line(0,1){0.12}}
\multiput(76.36,34.71)(0.11,0.13){3}{\line(0,1){0.13}}
\multiput(76.69,35.09)(0.11,0.13){3}{\line(0,1){0.13}}
\multiput(77,35.47)(0.1,0.13){3}{\line(0,1){0.13}}
\multiput(77.31,35.87)(0.15,0.2){2}{\line(0,1){0.2}}
\multiput(77.6,36.28)(0.14,0.21){2}{\line(0,1){0.21}}
\multiput(77.88,36.7)(0.13,0.21){2}{\line(0,1){0.21}}
\multiput(78.14,37.12)(0.13,0.22){2}{\line(0,1){0.22}}
\multiput(78.4,37.56)(0.12,0.22){2}{\line(0,1){0.22}}
\multiput(78.63,38)(0.11,0.22){2}{\line(0,1){0.22}}
\multiput(78.85,38.45)(0.1,0.23){2}{\line(0,1){0.23}}
\multiput(79.06,38.9)(0.1,0.23){2}{\line(0,1){0.23}}
\multiput(79.25,39.37)(0.18,0.47){1}{\line(0,1){0.47}}
\multiput(79.43,39.84)(0.16,0.47){1}{\line(0,1){0.47}}
\multiput(79.59,40.31)(0.15,0.48){1}{\line(0,1){0.48}}
\multiput(79.74,40.79)(0.13,0.48){1}{\line(0,1){0.48}}
\multiput(79.87,41.28)(0.11,0.49){1}{\line(0,1){0.49}}
\multiput(79.98,41.76)(0.1,0.49){1}{\line(0,1){0.49}}
\multiput(80.08,42.26)(0.08,0.49){1}{\line(0,1){0.49}}
\multiput(80.16,42.75)(0.07,0.5){1}{\line(0,1){0.5}}
\multiput(80.22,43.25)(0.05,0.5){1}{\line(0,1){0.5}}
\multiput(80.27,43.75)(0.03,0.5){1}{\line(0,1){0.5}}
\multiput(80.31,44.25)(0.02,0.5){1}{\line(0,1){0.5}}

\put(65,44.5){\makebox(0,0)[cc]{$\circ$}}

\put(72,52){\makebox(0,0)[cc]{$\circ$}}

\end{picture}

}

\def\figsix{

\def\JPicScale{0.8}
\ifx\JPicScale\undefined\def\JPicScale{1}\fi
\unitlength \JPicScale mm
\begin{picture}(80.33,60.33)(0,0)
\linethickness{0.3mm}
\put(80.32,44.75){\line(0,1){0.5}}
\multiput(80.31,45.75)(0.02,-0.5){1}{\line(0,-1){0.5}}
\multiput(80.27,46.25)(0.03,-0.5){1}{\line(0,-1){0.5}}
\multiput(80.22,46.75)(0.05,-0.5){1}{\line(0,-1){0.5}}
\multiput(80.16,47.25)(0.07,-0.5){1}{\line(0,-1){0.5}}
\multiput(80.08,47.74)(0.08,-0.49){1}{\line(0,-1){0.49}}
\multiput(79.98,48.24)(0.1,-0.49){1}{\line(0,-1){0.49}}
\multiput(79.87,48.72)(0.11,-0.49){1}{\line(0,-1){0.49}}
\multiput(79.74,49.21)(0.13,-0.48){1}{\line(0,-1){0.48}}
\multiput(79.59,49.69)(0.15,-0.48){1}{\line(0,-1){0.48}}
\multiput(79.43,50.16)(0.16,-0.47){1}{\line(0,-1){0.47}}
\multiput(79.25,50.63)(0.18,-0.47){1}{\line(0,-1){0.47}}
\multiput(79.06,51.1)(0.1,-0.23){2}{\line(0,-1){0.23}}
\multiput(78.85,51.55)(0.1,-0.23){2}{\line(0,-1){0.23}}
\multiput(78.63,52)(0.11,-0.22){2}{\line(0,-1){0.22}}
\multiput(78.4,52.44)(0.12,-0.22){2}{\line(0,-1){0.22}}
\multiput(78.14,52.88)(0.13,-0.22){2}{\line(0,-1){0.22}}
\multiput(77.88,53.3)(0.13,-0.21){2}{\line(0,-1){0.21}}
\multiput(77.6,53.72)(0.14,-0.21){2}{\line(0,-1){0.21}}
\multiput(77.31,54.13)(0.15,-0.2){2}{\line(0,-1){0.2}}
\multiput(77,54.53)(0.1,-0.13){3}{\line(0,-1){0.13}}
\multiput(76.69,54.91)(0.11,-0.13){3}{\line(0,-1){0.13}}
\multiput(76.36,55.29)(0.11,-0.13){3}{\line(0,-1){0.13}}
\multiput(76.01,55.66)(0.11,-0.12){3}{\line(0,-1){0.12}}
\multiput(75.66,56.01)(0.12,-0.12){3}{\line(0,-1){0.12}}
\multiput(75.29,56.36)(0.12,-0.11){3}{\line(1,0){0.12}}
\multiput(74.91,56.69)(0.13,-0.11){3}{\line(1,0){0.13}}
\multiput(74.53,57)(0.13,-0.11){3}{\line(1,0){0.13}}
\multiput(74.13,57.31)(0.13,-0.1){3}{\line(1,0){0.13}}
\multiput(73.72,57.6)(0.2,-0.15){2}{\line(1,0){0.2}}
\multiput(73.3,57.88)(0.21,-0.14){2}{\line(1,0){0.21}}
\multiput(72.88,58.14)(0.21,-0.13){2}{\line(1,0){0.21}}
\multiput(72.44,58.4)(0.22,-0.13){2}{\line(1,0){0.22}}
\multiput(72,58.63)(0.22,-0.12){2}{\line(1,0){0.22}}
\multiput(71.55,58.85)(0.22,-0.11){2}{\line(1,0){0.22}}
\multiput(71.1,59.06)(0.23,-0.1){2}{\line(1,0){0.23}}
\multiput(70.63,59.25)(0.23,-0.1){2}{\line(1,0){0.23}}
\multiput(70.16,59.43)(0.47,-0.18){1}{\line(1,0){0.47}}
\multiput(69.69,59.59)(0.47,-0.16){1}{\line(1,0){0.47}}
\multiput(69.21,59.74)(0.48,-0.15){1}{\line(1,0){0.48}}
\multiput(68.72,59.87)(0.48,-0.13){1}{\line(1,0){0.48}}
\multiput(68.24,59.98)(0.49,-0.11){1}{\line(1,0){0.49}}
\multiput(67.74,60.08)(0.49,-0.1){1}{\line(1,0){0.49}}
\multiput(67.25,60.16)(0.49,-0.08){1}{\line(1,0){0.49}}
\multiput(66.75,60.22)(0.5,-0.07){1}{\line(1,0){0.5}}
\multiput(66.25,60.27)(0.5,-0.05){1}{\line(1,0){0.5}}
\multiput(65.75,60.31)(0.5,-0.03){1}{\line(1,0){0.5}}
\multiput(65.25,60.32)(0.5,-0.02){1}{\line(1,0){0.5}}
\put(64.75,60.32){\line(1,0){0.5}}
\multiput(64.25,60.31)(0.5,0.02){1}{\line(1,0){0.5}}
\multiput(63.75,60.27)(0.5,0.03){1}{\line(1,0){0.5}}
\multiput(63.25,60.22)(0.5,0.05){1}{\line(1,0){0.5}}
\multiput(62.75,60.16)(0.5,0.07){1}{\line(1,0){0.5}}
\multiput(62.26,60.08)(0.49,0.08){1}{\line(1,0){0.49}}
\multiput(61.76,59.98)(0.49,0.1){1}{\line(1,0){0.49}}
\multiput(61.28,59.87)(0.49,0.11){1}{\line(1,0){0.49}}
\multiput(60.79,59.74)(0.48,0.13){1}{\line(1,0){0.48}}
\multiput(60.31,59.59)(0.48,0.15){1}{\line(1,0){0.48}}
\multiput(59.84,59.43)(0.47,0.16){1}{\line(1,0){0.47}}
\multiput(59.37,59.25)(0.47,0.18){1}{\line(1,0){0.47}}
\multiput(58.9,59.06)(0.23,0.1){2}{\line(1,0){0.23}}
\multiput(58.45,58.85)(0.23,0.1){2}{\line(1,0){0.23}}
\multiput(58,58.63)(0.22,0.11){2}{\line(1,0){0.22}}
\multiput(57.56,58.4)(0.22,0.12){2}{\line(1,0){0.22}}
\multiput(57.12,58.14)(0.22,0.13){2}{\line(1,0){0.22}}
\multiput(56.7,57.88)(0.21,0.13){2}{\line(1,0){0.21}}
\multiput(56.28,57.6)(0.21,0.14){2}{\line(1,0){0.21}}
\multiput(55.87,57.31)(0.2,0.15){2}{\line(1,0){0.2}}
\multiput(55.47,57)(0.13,0.1){3}{\line(1,0){0.13}}
\multiput(55.09,56.69)(0.13,0.11){3}{\line(1,0){0.13}}
\multiput(54.71,56.36)(0.13,0.11){3}{\line(1,0){0.13}}
\multiput(54.34,56.01)(0.12,0.11){3}{\line(1,0){0.12}}
\multiput(53.99,55.66)(0.12,0.12){3}{\line(1,0){0.12}}
\multiput(53.64,55.29)(0.11,0.12){3}{\line(0,1){0.12}}
\multiput(53.31,54.91)(0.11,0.13){3}{\line(0,1){0.13}}
\multiput(53,54.53)(0.11,0.13){3}{\line(0,1){0.13}}
\multiput(52.69,54.13)(0.1,0.13){3}{\line(0,1){0.13}}
\multiput(52.4,53.72)(0.15,0.2){2}{\line(0,1){0.2}}
\multiput(52.12,53.3)(0.14,0.21){2}{\line(0,1){0.21}}
\multiput(51.86,52.88)(0.13,0.21){2}{\line(0,1){0.21}}
\multiput(51.6,52.44)(0.13,0.22){2}{\line(0,1){0.22}}
\multiput(51.37,52)(0.12,0.22){2}{\line(0,1){0.22}}
\multiput(51.15,51.55)(0.11,0.22){2}{\line(0,1){0.22}}
\multiput(50.94,51.1)(0.1,0.23){2}{\line(0,1){0.23}}
\multiput(50.75,50.63)(0.1,0.23){2}{\line(0,1){0.23}}
\multiput(50.57,50.16)(0.18,0.47){1}{\line(0,1){0.47}}
\multiput(50.41,49.69)(0.16,0.47){1}{\line(0,1){0.47}}
\multiput(50.26,49.21)(0.15,0.48){1}{\line(0,1){0.48}}
\multiput(50.13,48.72)(0.13,0.48){1}{\line(0,1){0.48}}
\multiput(50.02,48.24)(0.11,0.49){1}{\line(0,1){0.49}}
\multiput(49.92,47.74)(0.1,0.49){1}{\line(0,1){0.49}}
\multiput(49.84,47.25)(0.08,0.49){1}{\line(0,1){0.49}}
\multiput(49.78,46.75)(0.07,0.5){1}{\line(0,1){0.5}}
\multiput(49.73,46.25)(0.05,0.5){1}{\line(0,1){0.5}}
\multiput(49.69,45.75)(0.03,0.5){1}{\line(0,1){0.5}}
\multiput(49.68,45.25)(0.02,0.5){1}{\line(0,1){0.5}}
\put(49.68,44.75){\line(0,1){0.5}}
\multiput(49.68,44.75)(0.02,-0.5){1}{\line(0,-1){0.5}}
\multiput(49.69,44.25)(0.03,-0.5){1}{\line(0,-1){0.5}}
\multiput(49.73,43.75)(0.05,-0.5){1}{\line(0,-1){0.5}}
\multiput(49.78,43.25)(0.07,-0.5){1}{\line(0,-1){0.5}}
\multiput(49.84,42.75)(0.08,-0.49){1}{\line(0,-1){0.49}}
\multiput(49.92,42.26)(0.1,-0.49){1}{\line(0,-1){0.49}}
\multiput(50.02,41.76)(0.11,-0.49){1}{\line(0,-1){0.49}}
\multiput(50.13,41.28)(0.13,-0.48){1}{\line(0,-1){0.48}}
\multiput(50.26,40.79)(0.15,-0.48){1}{\line(0,-1){0.48}}
\multiput(50.41,40.31)(0.16,-0.47){1}{\line(0,-1){0.47}}
\multiput(50.57,39.84)(0.18,-0.47){1}{\line(0,-1){0.47}}
\multiput(50.75,39.37)(0.1,-0.23){2}{\line(0,-1){0.23}}
\multiput(50.94,38.9)(0.1,-0.23){2}{\line(0,-1){0.23}}
\multiput(51.15,38.45)(0.11,-0.22){2}{\line(0,-1){0.22}}
\multiput(51.37,38)(0.12,-0.22){2}{\line(0,-1){0.22}}
\multiput(51.6,37.56)(0.13,-0.22){2}{\line(0,-1){0.22}}
\multiput(51.86,37.12)(0.13,-0.21){2}{\line(0,-1){0.21}}
\multiput(52.12,36.7)(0.14,-0.21){2}{\line(0,-1){0.21}}
\multiput(52.4,36.28)(0.15,-0.2){2}{\line(0,-1){0.2}}
\multiput(52.69,35.87)(0.1,-0.13){3}{\line(0,-1){0.13}}
\multiput(53,35.47)(0.11,-0.13){3}{\line(0,-1){0.13}}
\multiput(53.31,35.09)(0.11,-0.13){3}{\line(0,-1){0.13}}
\multiput(53.64,34.71)(0.11,-0.12){3}{\line(0,-1){0.12}}
\multiput(53.99,34.34)(0.12,-0.12){3}{\line(0,-1){0.12}}
\multiput(54.34,33.99)(0.12,-0.11){3}{\line(1,0){0.12}}
\multiput(54.71,33.64)(0.13,-0.11){3}{\line(1,0){0.13}}
\multiput(55.09,33.31)(0.13,-0.11){3}{\line(1,0){0.13}}
\multiput(55.47,33)(0.13,-0.1){3}{\line(1,0){0.13}}
\multiput(55.87,32.69)(0.2,-0.15){2}{\line(1,0){0.2}}
\multiput(56.28,32.4)(0.21,-0.14){2}{\line(1,0){0.21}}
\multiput(56.7,32.12)(0.21,-0.13){2}{\line(1,0){0.21}}
\multiput(57.12,31.86)(0.22,-0.13){2}{\line(1,0){0.22}}
\multiput(57.56,31.6)(0.22,-0.12){2}{\line(1,0){0.22}}
\multiput(58,31.37)(0.22,-0.11){2}{\line(1,0){0.22}}
\multiput(58.45,31.15)(0.23,-0.1){2}{\line(1,0){0.23}}
\multiput(58.9,30.94)(0.23,-0.1){2}{\line(1,0){0.23}}
\multiput(59.37,30.75)(0.47,-0.18){1}{\line(1,0){0.47}}
\multiput(59.84,30.57)(0.47,-0.16){1}{\line(1,0){0.47}}
\multiput(60.31,30.41)(0.48,-0.15){1}{\line(1,0){0.48}}
\multiput(60.79,30.26)(0.48,-0.13){1}{\line(1,0){0.48}}
\multiput(61.28,30.13)(0.49,-0.11){1}{\line(1,0){0.49}}
\multiput(61.76,30.02)(0.49,-0.1){1}{\line(1,0){0.49}}
\multiput(62.26,29.92)(0.49,-0.08){1}{\line(1,0){0.49}}
\multiput(62.75,29.84)(0.5,-0.07){1}{\line(1,0){0.5}}
\multiput(63.25,29.78)(0.5,-0.05){1}{\line(1,0){0.5}}
\multiput(63.75,29.73)(0.5,-0.03){1}{\line(1,0){0.5}}
\multiput(64.25,29.69)(0.5,-0.02){1}{\line(1,0){0.5}}
\put(64.75,29.68){\line(1,0){0.5}}
\multiput(65.25,29.68)(0.5,0.02){1}{\line(1,0){0.5}}
\multiput(65.75,29.69)(0.5,0.03){1}{\line(1,0){0.5}}
\multiput(66.25,29.73)(0.5,0.05){1}{\line(1,0){0.5}}
\multiput(66.75,29.78)(0.5,0.07){1}{\line(1,0){0.5}}
\multiput(67.25,29.84)(0.49,0.08){1}{\line(1,0){0.49}}
\multiput(67.74,29.92)(0.49,0.1){1}{\line(1,0){0.49}}
\multiput(68.24,30.02)(0.49,0.11){1}{\line(1,0){0.49}}
\multiput(68.72,30.13)(0.48,0.13){1}{\line(1,0){0.48}}
\multiput(69.21,30.26)(0.48,0.15){1}{\line(1,0){0.48}}
\multiput(69.69,30.41)(0.47,0.16){1}{\line(1,0){0.47}}
\multiput(70.16,30.57)(0.47,0.18){1}{\line(1,0){0.47}}
\multiput(70.63,30.75)(0.23,0.1){2}{\line(1,0){0.23}}
\multiput(71.1,30.94)(0.23,0.1){2}{\line(1,0){0.23}}
\multiput(71.55,31.15)(0.22,0.11){2}{\line(1,0){0.22}}
\multiput(72,31.37)(0.22,0.12){2}{\line(1,0){0.22}}
\multiput(72.44,31.6)(0.22,0.13){2}{\line(1,0){0.22}}
\multiput(72.88,31.86)(0.21,0.13){2}{\line(1,0){0.21}}
\multiput(73.3,32.12)(0.21,0.14){2}{\line(1,0){0.21}}
\multiput(73.72,32.4)(0.2,0.15){2}{\line(1,0){0.2}}
\multiput(74.13,32.69)(0.13,0.1){3}{\line(1,0){0.13}}
\multiput(74.53,33)(0.13,0.11){3}{\line(1,0){0.13}}
\multiput(74.91,33.31)(0.13,0.11){3}{\line(1,0){0.13}}
\multiput(75.29,33.64)(0.12,0.11){3}{\line(1,0){0.12}}
\multiput(75.66,33.99)(0.12,0.12){3}{\line(0,1){0.12}}
\multiput(76.01,34.34)(0.11,0.12){3}{\line(0,1){0.12}}
\multiput(76.36,34.71)(0.11,0.13){3}{\line(0,1){0.13}}
\multiput(76.69,35.09)(0.11,0.13){3}{\line(0,1){0.13}}
\multiput(77,35.47)(0.1,0.13){3}{\line(0,1){0.13}}
\multiput(77.31,35.87)(0.15,0.2){2}{\line(0,1){0.2}}
\multiput(77.6,36.28)(0.14,0.21){2}{\line(0,1){0.21}}
\multiput(77.88,36.7)(0.13,0.21){2}{\line(0,1){0.21}}
\multiput(78.14,37.12)(0.13,0.22){2}{\line(0,1){0.22}}
\multiput(78.4,37.56)(0.12,0.22){2}{\line(0,1){0.22}}
\multiput(78.63,38)(0.11,0.22){2}{\line(0,1){0.22}}
\multiput(78.85,38.45)(0.1,0.23){2}{\line(0,1){0.23}}
\multiput(79.06,38.9)(0.1,0.23){2}{\line(0,1){0.23}}
\multiput(79.25,39.37)(0.18,0.47){1}{\line(0,1){0.47}}
\multiput(79.43,39.84)(0.16,0.47){1}{\line(0,1){0.47}}
\multiput(79.59,40.31)(0.15,0.48){1}{\line(0,1){0.48}}
\multiput(79.74,40.79)(0.13,0.48){1}{\line(0,1){0.48}}
\multiput(79.87,41.28)(0.11,0.49){1}{\line(0,1){0.49}}
\multiput(79.98,41.76)(0.1,0.49){1}{\line(0,1){0.49}}
\multiput(80.08,42.26)(0.08,0.49){1}{\line(0,1){0.49}}
\multiput(80.16,42.75)(0.07,0.5){1}{\line(0,1){0.5}}
\multiput(80.22,43.25)(0.05,0.5){1}{\line(0,1){0.5}}
\multiput(80.27,43.75)(0.03,0.5){1}{\line(0,1){0.5}}
\multiput(80.31,44.25)(0.02,0.5){1}{\line(0,1){0.5}}

\put(65,44.5){\makebox(0,0)[cc]{$\circ$}}

\put(72,52){\makebox(0,0)[cc]{$\circ$}}

\put(72,58.5){\makebox(0,0)[cc]{$\star$}}

\end{picture}

}

\begin{document}

\baselineskip 24pt

\begin{center}

{\Large \bf Fixing an Ambiguity in Two Dimensional String Theory Using String Field Theory}

%{\Large \bf Classical Limit of Soft Theorems in Arbitrary Dimensions}

\end{center}

\vskip .6cm
\medskip

\vspace*{4.0ex}

\baselineskip=18pt

\centerline{\large \rm Ashoke Sen}

\vspace*{4.0ex}

\centerline{\large \it Harish-Chandra Research Institute, HBNI}
\centerline{\large \it  Chhatnag Road, Jhusi,
Allahabad 211019, India}

%\centerline{\large \it ~$^c$Homi Bhabha National Institute}
%\centerline{\large \it Training School Complex, Anushakti Nagar,
%    Mumbai 400085, India}

\vspace*{1.0ex}
\centerline{\small E-mail:  sen@hri.res.in}

\vspace*{5.0ex}

\centerline{\bf Abstract} \bigskip

In a recent paper, Balthazar, Rodriguez and Yin found some remarkable agreement 
between the results of c=1 matrix model and D-instanton corrections in two dimensional
string theory. Their analysis left undetermined two constants in the string  theory
computation which had to be fixed by comparing the results with the matrix model results.
One of these constants is affected by possible renormalization of the D-instanton action that
needs to be computed separately.   In this paper we fix the other constant by 
reformulating the world-sheet analysis in the language of string field theory.

\vfill \eject

\baselineskip 18pt

\tableofcontents

\section{Introduction} \label{s1}

In a recent paper\cite{1907.07688}, Balthazar, Rodriguez and Yin found some
remarkable agreement between the results of $c=1$ matrix model\cite{dj,sw,gk,kr} and two
dimensional string theory. In their analysis on the string theory side they left two
constants undetermined. Our goal in this paper will be to fix one of these constants
by carefully repeating their analysis in the language of string field theory. 
We find that the constant $c'$ in the analysis of \cite{1907.07688}, that was left undermined
in the world-sheet analysis, can be determined using the framework of string field theory and takes
value:
\be
c'=-\ln 4 \simeq -1.38629\, .
\ee
This is within $1\%$ of the numerical value $-1.399$ that was required for the agreement
between the results of world-sheet calculation and
the matrix model results.
The
other constant, called $S^{(1)}_{ZZ}$ in \cite{1907.07688},  is affected by 
possible renormalization of the D-instanton action that needs to be computed separately.
In particular since the D-instanton contribution to any
amplitude is proportional to $\exp[-c/g_s]$ for some positive constant $c$, a multiplicative 
renormalization of the D-instanton
action $c/g_s$ by $(1 -b \, g_s - a \, g_s^2)$ will generate an overall multiplicative factor of
$e^{c (b + a\, g_s)}$. Therefore contributions proportional to $g_s$
multiplying the leading order term will be affected by this renormalization of the D-instanton
action.\footnote{However 
once we have
determined the renormalization factor 
by comparing one amplitude, the same renormalization must continue to hold 
for all other amplitudes.}

We shall begin in \S\ref{s2} by describing the analysis of \cite{1907.07688} in the 
language of string field theory,
and the origin of the apparent ambiguity in their analysis. We shall then describe in \S\ref{s3} why
in a careful treatment using the framework of string field theory 
there is no ambiguity. Finally in \S\ref{s4} we shall use the machinery of
string field theory to fix the undetermined constant $c'$.

Before we turn to the details, let us first make a general remark on why string field theory is
useful for fixing the ambiguity described in \cite{1907.07688}. The ambiguity arises from having to
cancel the divergences between two diagrams, each of which is separately infrared divergent.
Therefore we need to regulate the diagrams before combining them, and since {\it a priori} the
regulators can be chosen independently, we can be left with a finite term after the cancellation that
depends on the finite parts of each regulated diagram. This is precisely the kind of situation where string
field theory is useful. String field theory makes clear the physical origin of the divergences and therefore
gives precise relation between the regulators in different diagrams, leaving behind no scope of an
ambiguity except those induced by field redefinition.

\sectiono{Formulation of the problem in the language of string field theory} \label{s2}

In this section we shall interpret the analysis of \cite{1907.07688} in the language of string field theory.
The main goal of \cite{1907.07688} was to compute the contribution of a single D-instanton to the
scattering of massless closed strings. Since the fluctuations of D-instanton are described by open
strings with ends on the D-instanton, the relevant string field theory is that of open and closed 
strings\cite{9705241}, with the open strings living on the 
D-instanton. Since our ultimate goal is to compute closed string scattering amplitudes,
it would be natural to integrate out the open string degrees of freedom and write an
effective action for closed string fields only. However on the D-instanton world-volume
there is a zero mode, describing the freedom of translating the D-instanton along the time
direction. This corresponds to a particular mode of the open string with vanishing propagator. 
For this reason, this mode cannot be integrated out at the beginning. The
best we can hope is to integrate out all the open string modes other than the zero
mode. The corresponding action still satisfies the 
BV master equation and was described as the Wilsonian effective action 
in \cite{1609.00459}.

Let us denote by $\phi$ the zero mode of the open string -- called $x^0$ in  \cite{1907.07688}. 
It was argued in \cite{1907.07688}  that the 
$\phi$ dependence of the Wilsonian action has a particularly simple form. 
The sum of the terms in the Wilsonian action with a fixed set of external closed string fields 
but arbitrary number of $\phi$ modes is proportional to
$\sum_n (i\omega\phi)^n/n!=e^{i\omega\phi}$ were $\omega$ is the total energy of all the
closed string fields. If we integrate over $\phi$ {\em after computing the Green's function of a set of
closed string fields in some background $\phi$}, we generate a factor proportional to
$\delta(\omega_{tot})$, -- $\omega_{tot}$ being the total energy of all the closed string states
in the Green's function. This restores energy conservation, which is otherwise broken for a fixed D-instanton
configuration. From this we can compute the S-matrix of closed string states by restricting the external
closed string states to be on-shell.

We are now in a position to describe the origin of the apparent ambiguity encountered in \cite{1907.07688} in the
language of string field theory.
The stems from the fact that the string field theory action is not unique. The construction of the action
requires the choice of local coordinates at the punctures where vertex operators are inserted, and
different choices lead to apparently different string field theories. As we shall elaborate in \S\ref{s3}, the apparent
ambiguity encountered in \cite{1907.07688} can be traced to this ambiguity. However, it follows from a general analysis
in \cite{9301097}, after suitable generalization to the field theory of open and closed strings, that the ambiguity described
above can be regarded as a result of field redefinition. Therefore it is not expected to change any of the physical
results. We shall see in \S\ref{s3}  how  the ambiguity in the result of  \cite{1907.07688} is actually resolved.

\section{String field theory as regulator} \label{s3}

We shall begin with some simple examples that illustrate how string field theory can be used to systematically 
analyze the infrared divergences in string theory, arising as ultraviolet divergences in the world-sheet 
theory\cite{1702.06489,1902.00263}.
Let us consider a unit disk with a bulk puncture at the centre and a boundary puncture at the point $z_1$
on the boundary of the disk. Using rotational symmetry we can fix $z_1$ to 1, but we shall keep it arbitrary for
later use. For defining off-shell amplitudes we need to choose local coordinates at both punctures,
with the property that the local coordinate vanishes at the puncture, and that the local coordinate around a boundary
puncture must take real values on the boundary. Let us for definiteness choose the local coordinate $w$ at the
bulk puncture to be $\beta z$ and the local coordinate $w_1$ at the boundary puncture to be 
\be \label{e0}
w_1 = i \, \lambda\, {(z_1-z)\over (z_1+z)}\, ,
\ee 
for
some real positive 
constants $\beta$ and $\lambda$. Note that \refb{e0} gives real $w_1$ as long as $z_1$ and $z$ are on the
boundary of the unit disk. It is convenient to take $\lambda$ and $\beta$ to be large -- in the language
of string field theory this is described as adding long stubs to the vertices\cite{9301097}. 
Once a choice of local coordinates is
made, we can compute a disk amplitude with an off-shell closed string vertex operator 
inserted at the center and an off-shell open
string vertex operator inserted at the boundary at the point $z_1$. 
This corresponds to an interaction vertex of string theory with
one open string and one closed string.

\begin{figure}
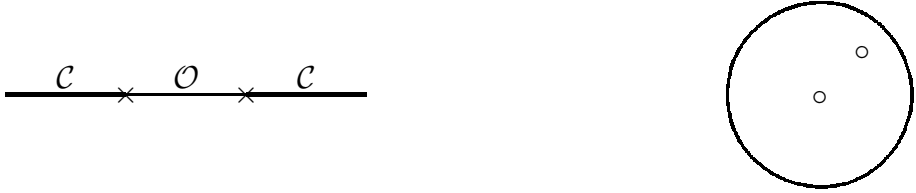


\begin{center}

\vskip -.5in

\hbox{\figone \quad \figfive}

\end{center}

\vskip -1.2in

\caption{The figure on the left shows a Feynman diagram obtained by joining the open string states from a pair
of open-closed interaction vertices by an open string propagator. The thick lines labelled by the symbol $\CC$
denote closed strings and the
thin line labelled by the symbol $\OO$ denotes open string.  A $\times$ denote an interaction vertex.
This is the convention we shall follow in all the subsequent figures. The figure on the right gives the 
representation of the Feynman diagram as a disk amplitude, with the closed string vertex operators, denoted by
$\circ$, inserted at 0 and $z_1(1-u)/(1+u)$.
\label{f1}}

\end{figure}

\begin{figure}
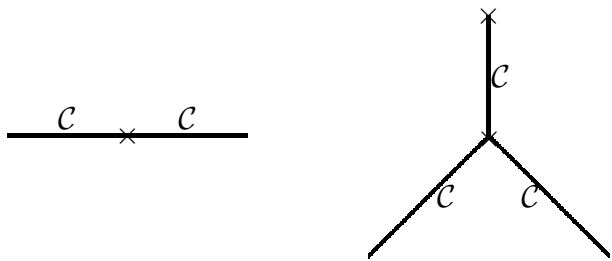


\begin{center}

\vskip -.5in

\figtwo

\end{center}

\vskip -1.5in

\caption{This diagram represents two other contributions, besides the one shown in Fig.~\ref{f1}, to the disk 
two point function of a pair of closed strings.
The first diagram represents
the two point closed string interaction vertex described by a disk amplitude with two closed strings, 
and the second diagram represents a Feynman diagram that joins the closed string three point vertex to a closed string one point vertex by a closed string propagator.  Since in our analysis the second diagram will not give a divergent
contribution, we shall include its contribution in the definition of the interaction vertex of two closed strings shown in the
first diagram.
\label{f2}}

\end{figure}

Now consider a Feynman diagram where we take two such vertices and join the open string states by a propagator.
This has been shown in Fig.~\ref{f1} with closed strings represented by thick lines and open strings represented by
thin lines.
This describes a disk amplitude with two external closed strings. However this does not cover the full moduli space
of the disk with two bulk punctures. Instead, it covers part of the moduli space that corresponds to the surface
\be\label{e1}
w_1 \tilde w_1 = -q, \quad 0\le q\le 1, \quad \Leftrightarrow \quad {(z-z_1) (\tilde z - z_1)\over (z+z_1) (\tilde z
+ z_1)} = q/\lambda^2 \equiv u\, , \quad 0\le u\le \eps, \quad \eps\equiv \lambda^{-2}\, ,
\ee
where $z$ and $\tilde z$ are the coordinates on two unit disks and $w_1=i\lambda (z_1-z)/(z_1+z)$, 
$\tilde w_1=i\lambda (z_1-\tilde z)/(z_1+\tilde z)$
are the local coordinates around the boundary punctures on the two disks. The identification \refb{e1} joins
the two disks into a single disk. If we parametrize this by the coordinate $z$, then the bulk punctures 
at $z=0$ and $\tilde z=0$ are situated
at 
\be \label{e2}
z=0, \qquad z=z_1(1-u)/(1+u)\, .
\ee 
Therefore as $u$ varies from $0$ to $\eps$ according to \refb{e1},
the location of the second puncture varies between $z_1$ and $z_1(1-\eps)/(1+\eps)$. For large
$\lambda$, this covers a small
part of the moduli space around the degenerate configuration where the second bulk puncture is close to the
boundary point $z_1$. The rest of the moduli space, where the second puncture lies on the line
segment between  $0$ and $z_1(1-\eps)/(1+\eps)$, needs to be covered by a combination of two other
Feynman diagrams
shown in Fig.~\ref{f2}.

In string field theory the parameter $q$ introduced in \refb{e1} appears as follows. An open string 
propagator contains a $1/L_0$ factor, which is
expressed as 
\be\label{e3}
(L_0)^{-1}=\int_0^1 dq \, q^{-1+L_0} = \eps^{-L_0}\, \int_0^\eps du \, u^{-1+L_0} \, , 
\ee
where $u$ and $\eps$ are the variables introduced in \refb{e1}. Upon substituting this into the expressions for the
Feynman diagram we can recover the geometric picture where connecting a pair of interaction vertices by a
propagator corresponds to sewing the surfaces via the relation \refb{e1}. 
Note however that \refb{e3} is valid only for states with positive $L_0$ eigenvalue. 
For $L_0< 0$, the right hand side diverges but the
left hand side is finite. Therefore while using the right hand side to find a geometric interpretation of the
amplitude is problematic, we can always use the left hand side. Equivalently, whenever we encounter
an expression of the form given in the right hand side of \refb{e3}, we use the replacement rule:
\be\label{e4}
\int_0^\eps du \, u^{-1+\alpha} \quad \Rightarrow \quad \eps^\alpha \, \alpha^{-1}\, .
\ee
Note that this does not work for $\alpha=0$. We shall return to this issue shortly. 

As discussed in \cite{1902.00263}, as
long as $\alpha$ is not an integer, the rule \refb{e4} is invariant under a change of variables $u\to v=a_1 u 
+a_2\, u^2+\cdots$
that is regular at $u=0$. Therefore given a divergent integral, we can apply this rule blindly without knowing if the
integration variable is actually the sewing parameter $u$ introduced in \refb{e1}. 
However for negative integer $\alpha$, application of this rule using a different variable $v$
could yield an expression that differs from the original expression by a constant. Therefore in this case we need to
identify the variable $u$ first before applying the rule.

In the case at hand, the intermediate open string in Fig.~\ref{f1} has a tachyonic mode that gives a contribution to the
integrand proportional to $u^{-2}$. The region of integration corresponding to $u>\eps$ comes from the
diagrams in Fig.~\ref{f2}, but the region of integration corresponding to $u\le\eps$ comes from Fig.~\ref{f1}.
Now suppose we want to integrate out the open string modes to construct the Wilsonian effective action for the
closed string fields. This would,
in particular, require
including the contribution from Fig.~\ref{f1} into the definition of the interaction vertex in the Wilsonian
effective action. For this we simply
add the contribution from Fig.~\ref{f1} after making the replacement \refb{e4} for the $\alpha=-1$ term representing
open string tachyon exchange. 
This procedure cannot be applied to the
$\alpha=0$ term, representing the effect of the zero mode exchange in Fig.~\ref{f1}.  This reflects the fact
that in constructing the Wilsonian effective action we cannot integrate out the open string
zero mode. We have to keep this 
unintegrated for now.

Next let us consider an interaction 
vertex with two external open strings and one external closed string, again associated with the
disk amplitude. Using appropriate SL(2,R) transformation we take the bulk puncture at the origin of the disk. 
The boundary
punctures then lie at points $z_1$ and $z_2$ on the disk. We now need to choose local coordinates 
at the punctures.
Let $w=\tilde\beta z$ describe the
local coordinate around the bulk puncture, and 
\be \label{e5}
w_a = i \tilde\lambda {(z_a-z)\over (z_a+z)}\, ,  \quad a=1,2\, ,
\ee 
be the local coordinates around the boundary punctures.  Here $\tilde\beta$ and 
$\tilde\lambda$ are some large positive
constants that are {\it a priori} independent of the constants $\beta$ and $\lambda$ appearing in \refb{e0}.
The choice \refb{e5}, although not unique, is consistent with
various symmetry requirements, {\it e.g.} symmetry under the cyclic permutation of the two open strings (which in this 
case is simply implemented by overall rotation of the $z$-plane). The choice of $\tilde\beta$ will not affect any
of our results since in our analysis the closed string inserted at the vertex will always be on-shell. 
The ambiguity mentioned in
\cite{1907.07688}  is related to the freedom of choosing $\tilde\lambda$ to be independent of
$\lambda$. As we shall now explain however, the choice $\tilde\lambda\ne\lambda$ is in conflict with another
hidden assumption that was used in the analysis of \cite{1907.07688}.

\begin{figure}
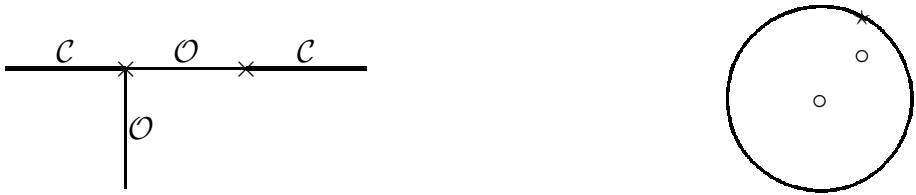


\begin{center}

\vskip -.6in

\hbox{\figthree \quad \figsix}

\end{center}

\vskip -1.2in

\caption{A Feynman diagram with two external closed strings and one external open string.
The right hand diagram is a representation of this as a disk amplitude, with the $\circ$'s denoting closed string vertex
operator and the $\star$ on the boundary denoting open string vertex operator.
\label{f3}}

\end{figure}

For this let us consider the result of joining the open-open-closed vertex to an open-closed vertex via an
open string propagator, as in Fig.~\ref{f3}. 
The Wilsonian effective action will contain contribution from this diagram, except the part
where the internal
open string state represents the zero mode. 
Now formally, using the representation \refb{e3} of the propagator,
the total contribution from Fig.~\ref{f3} can be
represented as an integral over certain
region of the moduli space of a disk with two closed string and one open string puncture.
In computing its contribution to the Wilsonian effective action, we need to express
the integrand as a sum over intermediate open string states, and subtract
the contribution of the internal $\phi$ mode 
from the integrand. 
Using analysis similar to those given in \refb{e1}, \refb{e2}, it is easy to see that the region of
the moduli space associated with Fig.~\ref{f3}
corresponds to one bulk puncture being located at the origin, the second bulk puncture
being located at $z_1(1-u)/(1+u)$ with $0\le u\le (\lambda\tilde \lambda)^{-1}$, and the boundary puncture
being
located at some point $z_2$.
Let us now take the external open string state to be the open string
zero mode $\phi$. 
Using the fact that the vertex operator of the  state $\phi$ is proportional to $\p X^0$ where 
$X^0$ is the world-sheet scalar field associated with the time coordinate, it is easy to see that 
after integration over $z_2$, the contribution
to the amplitude with an external open string zero mode 
field $\phi$ is given by $i\omega\phi$ times the amplitude where the
external open string state is removed.\footnote{When $z_2$ approaches $z_1$ in 
the open-open-closed vertex, there are potential
singularities and in the full string field theory we
represent the contribution from this region by another Feynman diagram
involving open-open-open vertex joined to an open-closed vertex. 
However for the term that we shall be analyzing, the contribution from this region will be
suppressed by positive powers of $\eps$.
This is reflected in the absence of singularity in the integrand in \refb{e8} 
in the $x\to 0$ limit and has been discussed in footnote \ref{fo5}.
Therefore for the analysis in this paper, we can include in the definition of open-open-closed vertex the
entire moduli space of the disk with two boundary punctures and one bulk puncture, 
including the region where the boundary
punctures are coincident. 
\label{fo2}
}
Here $\omega$ is the total energy carried by the external closed strings.
However over certain subspace of the moduli space, where the second bulk puncture is
located at $z_1(1-u)/(1+u)$ with $0\le u\le (\lambda\tilde \lambda)^{-1}$,  the amplitude will still
have missing internal $\phi$ state.
On the other hand the contribution to the term in the Wilsonian effective action without any external
$\phi$ insertion is given by a similar term, except that the subspace of the moduli space, over which the
internal $\phi$ contribution is removed, corresponds to the second bulk puncture being
located at $z_1(1-u)/(1+u)$ with $0\le u\le \lambda^{-2}$ (see \refb{e1}, \refb{e2}). 
If $\tilde\lambda=\lambda$, then these
two subspaces are identical, and we can express the sum of the terms with and without external $\phi$
as $(1+i\omega\phi)$ times the term without external $\phi$. This gives the first two terms in the
expansion of $e^{i\omega\phi}$. Requiring that the terms with higher powers of $\phi$ 
add up to the result $\exp[i\omega\phi]$ would demand {\em using the same parameter 
$\lambda$ in defining the local coordinates at the boundary punctures in the vertices with one closed and
multiple open strings}.

Since \cite{1907.07688} used $\exp[i\omega\phi]$ as the $\phi$ dependence of the Wilsonian effective action,
we see that we do not have the freedom of choosing $\tilde\lambda$ independently of $\lambda$ in
\refb{e5}. Instead we must choose:
\be\label{eequal}
\tilde\lambda=\lambda\, .
\ee

\begin{figure}
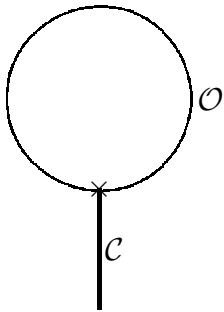


\begin{center}

\vskip -.7in

\hbox{\hskip 1.5in \figfour}

\end{center}

\vskip -.8in

\caption{A contribution to the annulus amplitude with one external closed string.
\label{f4}}

\end{figure}

We can now use this open-open-closed 
vertex to compute the contribution to the Wilsonian effective action with one external
closed string, obtained by joining the two open strings of the open-open-closed vertex by a propagator, as shown in
Fig.~\ref{f4}. 
Geometrically this describes an annulus amplitude with one external closed string. Naively, in order to
evaluate this contribution, the
$u$ parameter associated with the sewing
\be\label{e3.8}
w_1 w_2 = -q  \equiv -u\lambda^2\, ,
\ee
has to be integrated from 0 to $\eps$. However intermediate tachyons need special treatment, and 
we need to remove the contribution from
the state $\phi$ in the internal open string 
propagator. Following our earlier discussion, this leads to the following prescription
for computing the full Wilsonian effective action.
In the expression for the full annulus amplitude, expressed as integrals over the independent moduli 
parameters $u$ and $z_2/z_1$, we need to impose a 
lower cut-off $\eps=\lambda^{-2}$ on the $u$ integral and add a compensating term
using the replacement
rule \refb{e4} for $\alpha=1$. This would correspond to integrating out the tachyonic mode in Fig.~\ref{f4}, but
leaving the zero mode unintegrated.
Integration over the zero mode will have to be taken care of at the end.

To summarize, we need to augment the world-sheet analysis of \cite{1907.07688} using the following
procedure for regulating divergences:
\begin{enumerate}
\item When we encounter a divergence, we need to identify the degeneration responsible for the divergence,
and change the integration variables to the analog of the parameter $u$ appearing in sewing relations of type
described in \refb{e1} and the moduli of the Riemann surfaces that are being sewed.
\item If near $u=0$ the integrand has a term of the form $A\, du\, u^{-2}$ for some $u$ independent $A$, 
we replace the lower limit
of integration by $\eps$, and subtract $A\, \eps^{-1}$ from the integral. The subtraction of $A\, \eps^{-1}$
represents the effect of `integrating out' the tachyon field in the Wilsonian effective action.
\item If near $u=0$ the integrand has a term of the form $B\, du\, u^{-1}$ for some $u$ independent $B$, we
replace the lower limit of $u$ integration by $\eps$ without adding any compensating term. A 
term proportional to $du\, u^{-1}$ in the integrand represent the effect of zero mode exchange, and
we cannot integrate out these modes. 
\item Integration over the open string zero modes is performed at the end {\em after computing the
desired Green's function / S-matrix}. This produces the energy
conserving delta function.
\end{enumerate}

\sectiono{Cancellation of the extra terms} \label{s4}

In this section we shall implement the general principles described in \S\ref{s3} and determine the
constant that was left unfixed in the world-sheet analysis of \cite{1907.07688}. We shall begin by
describing the strategy that we shall follow.

The analysis of \cite{1907.07688} involved addition of some extra terms to the result of the world-sheet
computation on the string theory side. These terms were used to regulate the infrared divergences,
and are given in the second, fourth and fifth lines of
eq.(2.36) of \cite{1907.07688}. They take the following form:
\ben\label{e6}
&& e^{-1/g_s}\, g_s \, \delta(\omega-\omega') \, 8\, \pi\, \NN\,
\Bigg[-{A\over 16} \, \pi^{1/2} 2^{-1/4} \,
\left(\Psi^{ZZ}(\omega/2)\right)^2 \int_0^1 dy\, y^{-2} (1-2\, \omega^2 y) \nonumber \\ && 
\hskip 1in -{A\over 4}\, 2^{3/4}\,
\pi^{3/2} \left(\Psi^{ZZ}(\omega/2)\right)^2 \int_0^\infty dt \, \int_0^{1/4} dx \, 
\left( {e^{2\pi t} -1\over \sin^2 (2\pi x)} + 2\omega^2\right) \nonumber \\ &&\hskip 1in
- S^{(1)}_{ZZ} \, \sinh^2(\pi\omega) + c' \omega^2 \sinh^2\pi\omega
\Bigg]\, ,
\een
where $\NN$ is a normalization constant,
\be
\Psi^{ZZ} (\omega/2) = 2^{5/4}\, \sqrt \pi\, \sinh(\pi\omega)\, , \quad A= 2^{3/4}\, \pi^{-3/2}\, ,
\ee
and $S^{(1)}_{ZZ}$ and $c'$ are constants which were eventually adjusted to make the amplitude agree
with the matrix model result. Since our goal will be to show that string theory results agree with the matrix
model results without any ad hoc adjustment, we need to show that \refb{e6} vanishes after we use the
regulator implied by string field theory.\footnote{The logic goes as follows. Since the counterterms \refb{e6}
have been chosen to cancel the divergences in the world-sheet result, if we regulate both the original world-sheet
results and the counterterms following the prescription given by string field theory, the result remains unchanged,
But now the regulated world-sheet integrals give the correct result. Therefore the regulated counterterms must add
up to zero.}

The terms inside the square bracket in \refb{e6} can be divided into two classes: the ones proportional to
$\sinh^2(\pi\omega)$ and the ones proportional to $\omega^2 \sinh^2(\pi\omega)$. The leading term (not
displayed here) is also proportional to $\sinh^2(\pi\omega)$. Therefore, renormalization of the 
D-instanton action,
mentioned in the first paragraph of the introduction, will change the coefficient of 
the term proportional to $\sinh^2(\pi\omega)$. The coefficient of the $\omega^2\sinh^2(\pi\omega)$ term cannot
be changed this way, and so we focus on that term. Changing the integration variable $t$ to 
\be
v=e^{-2\pi t}\, ,
\ee 
we can
express the terms proportional to $\omega^2\sinh^2(\pi\omega)$ inside the square bracket as:
\be \label{e8}
\left\{\int_0^1 dy\, y^{-1} - 4\, \int_0^{1/4} dx \, \int_0^1 dv\, v^{-1} +c'
\right\} \, \omega^2\sinh^2(\pi\omega)\, .
\ee
The first integral is divergent at $y=0$ while the second integral is divergent at $v=0$. Our goal now will be to
identify the source of these divergences to appropriate degenerations of  Riemann surfaces, find the relation
between the integration variables in \refb{e8} and the sewing parameter $u$ in \refb{e1} or 
\refb{e3.8} and possibly other
moduli of the Riemann surfaces, and then translate the cut-off
procedure on $u$ prescribed by string field theory to a cut-off on $y$ and $v$.

We begin with the $y$ integral in the first term in \refb{e8}. 
This term was chosen to remove the divergences in the two point function on the
disk, with the closed string vertex operators inserted at $i$ and $i\, y$ in the upper half plane\cite{1907.07688},
which we shall label by $w$.
If we denote by $z$ the coordinate on the unit disk, related to $w$ via 
$z = i {i-w\over i+w}$, then in the $z$ coordinate the punctures are at $z=0$ and $z=i (1-y)/(1+y)$. Therefore
the $y\to 0$ limit corresponds to one of the closed string vertex operators coming close to the boundary point
$z=i$. Comparing this with \refb{e2} we see that this corresponds to the degeneration of an open string propagator
joining two disks associated with the Feynman diagrams shown in Fig.~\ref{f1}, with the identification:
\be
z_1=i, \quad y = u\, .
\ee
Therefore the cut-off $u>\eps$ translates to $y>\eps$.

Next we turn to the divergence of the second term in \refb{e8} in the $v\to 0$ limit. 
In the analysis of \cite{1907.07688} this comes from the one point function of a
closed string state on the annulus, parametrized by coordinate $w$, satisfying,
\be\label{ep1}
0\le Re(w)\le \pi, \quad w \sim w + 2\pi i t = w - i\, \ln v\, ,
\ee
with the bulk puncture located at $w=2\pi x$. Here $\sim$ denotes identification of points under the given
transformation. Our goal will be to compare this configuration with the one
associated with the diagram shown in Fig.~\ref{f4}. The latter is parametrized by  the sewing parameter $u$ 
appearing in \refb{e3.8}, and the ratio $z_2/z_1$,
with $z_1$, $z_2$ labelling the locations of the boundary punctures on the disk in the open-open-closed interaction
vertex. Once we have determined the relation between
$(x,v)$ and $(z_2/z_1, u)$, we can translate the cut-off $u>\eps$ 
into a cut-off on the $(x,v)$ integration and carry
out the integration over $x$ and $v$.

We begin by recalling the choice of local coordinates around the two boundary punctures in the open-open-closed
interaction vertex:\footnote{If we had taken $\tilde\lambda$ in \refb{e5} to be different from $\lambda$, the
$\lambda$ in \refb{e5a} will change to $\tilde\lambda$. Subsequent equations will change accordingly, 
replacing $\eps$ by $\tilde\eps\equiv\tilde\lambda^{-2}$
in \refb{et22}.}
\be \label{e5a}
w_a = i \, \lambda {(z_a-z)\over (z_a+z)}\, ,  \quad a=1,2\, ,
\ee 
and the sewing relation\be\label{es1}
w_1 \, w_2 = -q = -\lambda^2 \, u \quad \Leftrightarrow \quad {(z_1-z)\over (z_1+z)} \sim u
\, {(z_2+z)\over (z_2-z)}\, .
\ee
We now introduce the coordinate 
\be \label{es1.5}
\tilde w = i {z_1-z\over z_1+z} \, ,
\ee
taking value in the upper half plane, 
and define
\be
e^{2i\theta} \equiv {z_2\over z_1}\, .
\ee
In the $\tilde w$ coordinate the sewing relation \refb{es1} takes the form
\be \label{es2}
\tilde w \sim {u^{1/2}\,  \sin\theta \, \tilde w + u^{1/ 2} \, \cos\theta\over -u^{-1/2} 
 \cos\theta\, \tilde w + u^{-1/2} \sin\theta}\, .
\ee
For small $u$, the right hand side of \refb{es2} is an SL(2,R) transformation with a hyperbolic element. We can 
diagonalize this by an SL(2,R) transformation $\pmatrix{a & b\cr c & d}$ such that
\be
\pmatrix{a & b\cr c & d}^{-1}  
\pmatrix{u^{1/2}\,  \sin\theta & u^{1/ 2} \, \cos\theta\cr -u^{-1/2} 
 \cos\theta & u^{-1/2} \sin\theta}
\pmatrix{a & b\cr c & d} = \pmatrix{\beta^{1/2} & 0\cr 0 & \beta^{-1/2}}\, ,
\ee
so that in the new coordinate system $\hat w$, related to $\tilde w$ via
\be
\tilde w = {a\, \hat w+b\over c\, \hat w+d}\, ,
\ee
the identification \refb{es2} takes the form
\be\label{es3}
\hat w \sim \beta \, \hat w\, .
\ee
We shall not give the explicit form of $\beta$ and $\pmatrix{a & b\cr c & d}$ for general $u$, since we only
need this for small $u$. In the small $u$ limit,
\be\label{es4}
\beta = u^{-1} \, \sin^2\theta, \quad \pmatrix{a & b\cr c & d} =\pmatrix{0 & 1\cr -1 & \cot\theta}\, .
\ee
This gives 
\be  \label{es5}
\tilde w = {1\over -\hat w + \cot\theta}, \quad \hat w = {\cot\theta\, \tilde w - 1\over \tilde w}\, .
\ee
Finally we map the upper half plane spanned by $\hat w$ to the strip spanned by the coordinate $w$,
defined as:
\be
w = {1\over i} \, \ln\hat w\, .
\ee
Since $\hat w$ takes value in the upper half plane, we get
\be 
0\le {\rm Re}(w) \le \pi\, .
\ee
Furthermore the identification \refb{es3} with $\beta$ defined in \refb{es4} gives
\be
w \sim w + i\ln \beta = w + i\ln (u^{-1} \sin^2\theta)\, .
\ee
Comparing this with \refb{ep1} we get
\be \label{et1}
v = u / \sin^2\theta\, .
\ee

It remains to locate the position of the bulk puncture in the $w$ plane since its real part
gives the value of
$2\pi x$, -- due to translational invariance of the annulas, ${\rm Im}\, w$ does not carry any
physical information. Since in the $z$-plane the bulk puncture 
is located at $z=0$, we see from \refb{es1.5} that in the $\tilde w$ plane
it is at $\tilde w=i$. \refb{es5} now gives its location at $\hat w=\cot\theta+i = e^{i\theta}/\sin\theta$ and
therefore at $w = \theta + i \ln \sin\theta$. Since $2\pi x$ is the real part of $w$, we get
\be\label{et2}
2\, \pi\, x = \theta\, .
\ee
\refb{et1} and \refb{et2} now shows that the cut-off $u\ge\eps$ can be translated 
to\footnote{Eq.\refb{et22} is valid only for $\sin^2(2\pi x) >> \eps$ since in arriving
at \refb{et1} we have used the approximation $u<<\sin^2\theta$. Furthermore \refb{es2} ceases to represent
a hyperbolic element of SL(2,R) for $(u^{1/2}+u^{-1/2})\sin\theta \le 1$. This is related to the issues discussed in
footnote \ref{fo2}. However since the contribution to the integral from the region  $\sin^2(2\pi x) \sim \eps$
is suppressed by powers of $\eps$, we ignore this complication. \label{fo5}} 
\be\label{et22}
v\ge \eps/\sin^2(2\pi x)\, .
\ee

Using \refb{et22}, and the earlier result that the cut-off on the $y$ integral is $y\ge \eps$, we can express
\refb{e8} as
\ben \label{e88}
&& \left\{\int_\eps^1 dy\, y^{-1} - 4\, \int_0^{1/4} dx \, \int_{\eps/\sin^2(2\pi x)}^1 dv\, v^{-1} +c'
\right\} \, \omega^2\sinh^2(\pi\omega) 
\nonumber \\
&=& \left(c'-4\, \int_0^{1/4} dx \, \ln(\sin^2(2\pi x))\right)\, \omega^2\sinh^2(\pi\omega)\nonumber \\ 
&=&  \left(c'+\ln 4\right) \, \omega^2\sinh^2(\pi\omega)\, .
\een
Demanding that this ad hoc term vanishes, gives
\be 
c'=-\ln 4 \simeq -1.38629\, .
\ee
This is within $1\%$ of the numerical value $-1.399$ determined in \cite{1907.07688} by 
comparing the string theory
results with the matrix model results. 

\bigskip

{\bf Acknowledgement:}
We would like to thank Bruno Balthazar, Victor Rodriguez,
Xi Yin and Barton Zwiebach for useful communications and comments on an earlier  draft of the paper.
The work of A.S. was
supported in part by the 
J. C. Bose fellowship of 
the Department of Science and Technology, India and the Infosys chair professorship.

\end{document}